\documentclass[twocolumn,showpacs,preprintnumbers]{revtex4}
\usepackage{amssymb}
\usepackage{amsmath}
\usepackage{epsfig}
\usepackage{graphics}

\begin{document}

\title{Spin injection across magnetic/non-magnetic interfaces with finite
magnetic layers}
\author{Alexander Khaetskii$^{1}$}
\author{J. Carlos Egues$^{1,2}$}
\author{Daniel Loss$^{1}$}
\affiliation{$^1$Department of Physics and Astronomy, University
of Basel, Klingelbergstrasse 82, CH-4056 Basel, Switzerland}
\affiliation{$^2$Departamento de F\'{\i}sica e Inform\'{a}tica,
Instituto de F\'{\i}sica de S\~{a}o Carlos, Universidade de
S\~{a}o Paulo, 13560-970 S\~{a}o Carlos, S\~{a}o Paulo, Brazil}
\author{Charles Gould, Georg Schmidt, and Laurens W. Molenkamp}
\affiliation{Physikalisches Institut, Universit\"{a}t W\"{u}rzburg, Am Hubland, 97074 W%
\"{u}rzburg, Germany}
\date{\today}

\begin{abstract}
We have reconsidered the problem of spin injection across
ferromagnet/non-magnetic-semiconductor (FM/NMS) and
dilute-magnetic-semiconductor/non-magnetic-semiconductor
interfaces, for structures with \textit{finite} magnetic layers
(FM or DMS). By using appropriate physical boundary conditions, we
find expressions for the resistances of these structures which are
in general different from previous results in the literature. When
the magnetoresistance of the contacts is negligible, we find that
the spin-accumulation effect alone cannot account for the $d$
dependence observed in recent magnetoresistance data. In a limited
parameter range, our formulas predict a strong $d$ dependence
arising from the magnetic contacts in systems where their
magnetoresistances are sizable.

\end{abstract}

\pacs{75.50.Pp, 73.50.Jt, 73.61.Ga}

\maketitle

\section{Introduction}

Spin injection across interfaces is one of the crucial ingredients for the
successful implementation of novel spintronic devices \cite{wolf,als,Schmidt}%
. For instance, the spin-transistor proposal \cite{datta} relies on
subjecting injected spin-polarized electrons to a controllable spin
precession between the ferromagnetic source and drain. Hybrid
ferromagnetic/semiconductor and dilute-magnetic-semicondutor/non-magnetic
semiconductor junctions constitute basic interfaces in which to investigate
spin-polarized transport. However, the efficiency of spin-injection through
ideal FM/NMS interfaces turns out to be disappointingly small due to the
large conductivity mismatch \cite{Schmidt1} between the metallic ferromagnet
and the semiconductor. The use of spin-selective interfaces can
significantly enhance injection efficiencies \cite{rashba}. This can be
accomplished by inserting spin-dependent tunnel barriers between the FM and
the NMS layers \cite{motsnyi}.

Particularly promising is spin injection from a dilute-magnetic
semiconductor (DMS) into a non-magnetic semiconductor (NMS). Novel DMS/NMS
junctions (i) minimize the conductivity mismatch problem and (ii) naturally
incorporate spin dependency in the transmission process \cite{egues}. As
recently demonstrated, substantial spin injection can be achieved in these
Mn-based heterostructures \cite{spin-pol}. More recently, a novel large
magnetoresistance effect has been observed in a DMS/NMS/DMS geometries \cite%
{mol}. An interesting observation of Ref. \cite{mol} is the dependence of
the magnetoresistance effect on the thickness of the DMS layers: the MR
effect doubles with the DMS thickness.

Available formulas describing GMR-type effects in
magnetic/non-magnetic junctions assume 1D geometries with
semi-infinite magnetic layers \cite{Khaetskii}. In this case, the
expressions for the electrochemical potentials for the spin-up and
the spin-down electrons contain only decaying exponentials in the
magnetic regions \cite{formulas}. The resistances in the parallel
and antiparallel configurations are calculated assuming that the
device extends for a spin-flip length into the magnetic contacts.
In such an approach it is not clear how to properly take into
account the voltage drop across the sample.

\begin{figure}[th]
\begin{center}
\epsfig{file=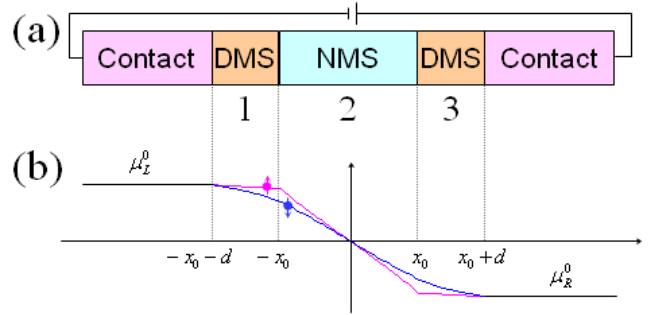,width=0.5\textwidth}
\end{center}
\caption{Color online. Schematic of the magnetic/non-magnetic structure
investigated here (a) and the corresponding spin-resolved
electrochemical potentials (b) versus the position across the
layers. Note that we consider magnetic layers (DMS) of finite
widths \emph{d} symmetrically placed around the non-magnetic layer
(NMS). The spin-splitting of the electrochemical potentials across
the magnetic/non-magnetic interfaces is larger for a higher
\textquotedblleft conductivity mismatch\textquotedblright\ between
these layers.} \label{fig:fig1}
\end{figure}
In this work we consider DMS/NMS/DMS and FM/NMS/FM 1D systems with
\emph{finite} magnetic layers, Fig. 1. As detailed below, we
describe diffusive transport in these structures via the usual
diffusion theory of van Son et al. \cite{vanson} with proper
physical boundary conditions between the several
magnetic/non-magnetic interfaces in the system. The formulas for
the magnetoresistance obtained within our treatment are in general
different from the ones previously obtained in literature, even in
the experimentally relevant regime $d \gg \lambda _{D}$ and
$x_{0}\leq \lambda _{N}$; $\lambda _{D},\lambda _{N}$ denote the
spin-flip lengths in the DMS and NMS layers, respectively, $d$ is
the length of each DMS layer and $2x_{0}$ is the length of the NMS
layer. However, in the case of a FM/NMS/FM system where the
conductivity mismatch is large, this difference is not important
since the correction we find is very small in this case. For
DMS/NMS/DMS systems, on the other hand, the conductivity mismatch
is small and the magnetoresistance effect can be large. Our
formulas significantly deviate from the earlier ones in this case,
especially in the regime where $\lambda _{D}$, which can be
magnetic-field dependent, is comparable to $d$.

\section{System and approach}

Let us denote by 1,2, and 3 the regions corresponding to the DMS
(or FM), NMS, DMS (or FM) layers, respectively. Note that in Fig.
1 $x=0$ corresponds to the center of the structure (NMS layer),
$x=\pm x_{0}$ correspond to the DMS/NMS interfaces and $x=\pm
(x_{0}+d)$ to the metal contact/DMS interfaces. Consider first the
DMS/NMS/DMS case or, equivalently, a parallel configuration
FM/NMS/FM. Our task is to find the resistance of the structure
$R=V/j$, where $j=j_{\uparrow }+j_{\downarrow }$ is the total
current through the structure and $V$ is an applied bias ($+V/2$
is the potential of the left metal contact and $-V/2$ is the
potential of the right contact).

\textit{Diffusive transport equations.} Our starting point is the
diffusive transport approach of\ Ref. \onlinecite{vanson}, with
the basic equations relating the spin-dependent electrochemical
potentials and current densities
\begin{equation}
j_{\uparrow ,\downarrow }=(\sigma _{\uparrow ,\downarrow }/e)d\mu _{\uparrow
,\downarrow }/dx.  \label{eq2}
\end{equation}
which hold in all three regions 1, 2, and 3 (see Fig. 1) of our
system; $\sigma _{\uparrow ,\downarrow }$ denotes the conductivity
of the spin up (or down) electrons in the respective layer. The
difference in the spin-dependent electrochemical potential is
\cite{vanson}
\begin{equation}
\mu_\uparrow -\mu_\downarrow = \lambda^2 \frac{d^2(\mu_\uparrow
-\mu_\downarrow )}{dx^2}, \label{eq2a}
\end{equation}
where $\lambda$ is the spin-flip length ($\lambda_D$ or
$\lambda_N$) in the corresponding layer. As usual, to solve the
problem we first assume a general solution in each of the three
regions and then match them at the many interfaces via proper
boundary conditions. Below we outline this procedure, clearly
stating the physical boundary conditions we use and the
simplifying symmetries of the problem.

\emph{General solution for }$\mu _{\uparrow ,\downarrow }$\emph{.
} Consistent with Eq. (\ref{eq2a}), we use the general expression
for the electrochemical potentials in region 1 (DMS)
\begin{widetext}
\begin{equation}
  \frac{1}{-e}  \left( \begin{array}{c}
\mu_{\uparrow}\\ \mu_{\downarrow}
\end{array}
\right)   = (A_1 x + B_1)\left( \begin{array}{c}
1\\1 \end{array}
\right)    + (C_1  \lambda _{D} \exp(x/\lambda _{D}) + D_1  \lambda _{D} \exp(-x/\lambda _{D}))
 \left( \begin{array}{c}
1/\sigma_{\uparrow}^D\\ -1/\sigma_{\downarrow}^D
\end{array}
\right). \label{potent}
\end{equation}
\end{widetext}Here $\lambda=\lambda _{D}$ is the spin flip length in the DMS layer.
Note the presence of the exponentially-growing term in Eq.
(\ref{potent}), since we consider a \textit{finite} DMS layer. The
corresponding expressions for the electrochemical potentials in
the regions 2 and 3 are obtained from Eq. (\ref{potent}) by the
replacements $A_{1}\rightarrow A_{2,3}$, $B_{1}\rightarrow
B_{2,3}$, $C_{1}\rightarrow C_{2,3}$, and $D_{1}\rightarrow
D_{2,3}$; in addition we should make $\sigma _{\uparrow
,\downarrow }^{D}\rightarrow \sigma =\sigma _{N}/2$ and $\lambda
_{D}\rightarrow \lambda _{N}$ in the NMS region 2. The chosen form
of the solution guarantees conservation of the total current
everywhere in the sample. From the symmetry of the problem in the
parallel configuration considered  here, it follows that the
electrochemical potentials $\mu _{\uparrow }$ and $\mu
_{\downarrow }$ should be odd functions of the coordinate $x$, see
Fig. 1(b). In terms of the coefficients this means that $%
A_{3}=A_{1},B_{3}=-B_{1},C_{3}=-D_{1},D_{3}=-C_{1},B_{2}=0$, and $%
C_{2}=-D_{2}$. Hence we can consider only the boundary conditions at the
left metal contact/DMS interface and left DMS/NMS interface.

\emph{Boundary conditions.} At the metal contact/DMS interface we use the
continuity of the \emph{total} current and the equality of the
electrochemical potentials: $\mu _{\uparrow }=\mu _{\downarrow }=-eV/2$. The
condition $\mu _{\uparrow }=\mu _{\downarrow }$ is consistent with assuming
a metal contact with infinite conductivity.  Thus we
obtain the following set of equations: 
\begin{equation}
C_{1}+D_{1}\exp (2(x_{0}+d)/\lambda _{D})=0,  \label{boundary1}
\end{equation}%
\begin{equation}
-A_{1}(x_{0}+d)+B_{1}=V/2,  \label{boundary2}
\end{equation}%
and
\begin{equation}
j=-A_{1}(\sigma _{\uparrow }^{D}+\sigma _{\downarrow }^{D}).
\label{boundary3}
\end{equation}

At the DMS/NMS interface we use the continuity of the electrochemical
potentials, $\mu _{\uparrow }(-x_{0}^{-})=\mu _{\uparrow }(-x_{0}^{+})$, $
\mu _{\downarrow }(-x_{0}^{-})=\mu _{\downarrow }(-x_{0}^{+})$, the
conservation of the total current $j=j_{\uparrow }+j_{\downarrow }$, and
conservation of the $j_{\uparrow }$ component across the interface. Then we
have the following additional set of equations
\begin{widetext}
\begin{eqnarray}
-A_1 x_0 + B_1 + \left( \begin{array}{c}
\lambda _{D}/\sigma_{\uparrow}^D\\ -\lambda _{D}/\sigma_{\downarrow}^D
 \end{array}
\right)(C_1   \exp(-x_0/\lambda _{D}) + D_1 \exp(x_0/\lambda _{D})) =
-A_2x_0 \mp 2C_2\lambda_N \sinh (x_0/\lambda_N )/\sigma,  \label{boundary4} \\
-A_1 \sigma_{\uparrow }^{D}-(C_1\exp(-x_0/\lambda _{D}) -D_1
\exp(x_0/\lambda _{D}))= -A_2\sigma -2C_2 \cosh(x_0/\lambda_N),
\label{boundary5}
\end{eqnarray}
\end{widetext}and
\begin{equation}
j=-2\sigma A_{2}.  \label{boundary6}
\end{equation}
Hence we have seven equations, Eqs.
(\ref{boundary1})--(\ref{boundary6}), for seven unknown quantities
$A_{1},A_{2},B_{1},C_{1},C_{2},D_{1}$, and $j$. By solving these
equations we can determine the resistance of the structure. This
we do next.

\section{Results}

For a FM/NMS/FM and a DMS/NMS/DMS system (\textquotedblleft
parallel\textquotedblright\ configuration), we find for
$R_{p}=V/j$
\begin{eqnarray}
R_{p} &=&\frac{2x_{0}}{\sigma ^{N}}+\frac{2d}{\sigma ^{D}}+  \notag \\
&&\frac{2\beta ^{2}}{(1-\beta ^{2})\frac{\sigma ^{D}}{\lambda
_{D}}\coth (\frac{d}{\lambda_{D}})+\frac{\sigma ^{N}}{\lambda
_{N}}\coth (\frac{x_{0}}{\lambda_{N}})}, \label{resist}
\end{eqnarray}
where $\beta =(\sigma _{\uparrow }^{D}-\sigma _{\downarrow
}^{D})/(\sigma _{\uparrow }^{D}+\sigma _{\downarrow }^{D})$,
$\sigma ^{N}=2\sigma $ is the total conductivity in the NMS layer
and $\sigma ^{D}=\sigma _{\uparrow }^{D}+\sigma _{\downarrow
}^{D}$ is the conductivity in the DMS layer; $d$ is the length of
each DMS layer and $2x_{0}$ is the length of the NMS layer. The
term proportional to the magnetic-layer width $d$ in Eq.
(\ref{resist}) is not present in earlier formulas in the
literature. Note that the term proportional to $\beta ^{2}$ in Eq.
(\ref{resist}) differs from that in Eq. (1) of Ref. \cite{mol}
because it is $d$ dependent and also because of its distinctive
$x_0$ dependence \cite{correction}.

For the antiparallel configuration (opposite FM layers), we can
similarly obtain an expression for the resistance $R_{ap}$.
Interestingly, $R_{ap}$ can be obtained from $R_p$ in
Eq.(\ref{resist}) by changing $\tanh (x_{0}/\lambda _{N})$ to
$\coth (x_{0}/\lambda _{N})$
\begin{eqnarray}
R_{ap} &=&\frac{2x_{0}}{\sigma ^{N}}+\frac{2d}{\sigma ^{D}}+  \notag \\
&&\frac{2\beta ^{2}}{(1-\beta ^{2})\frac{\sigma ^{D}}{\lambda _{D}}%
\coth (\frac{d}{\lambda _{D}})+\frac{\sigma ^{N}}{\lambda_N}\tanh
(\frac{x_{0}}{\lambda_N})}.  \label{anti}
\end{eqnarray}
In obtaining Eq. (\ref{anti}) we have also assumed that the
spin-up and the spin-down conductivities of region 3 are
$\sigma_\downarrow^D$ and $\sigma_\uparrow^D$, respectively. We
emphasize that the term proportional to $\beta^2$ [Eqs.
(\ref{resist}), (\ref{anti})] does not reduce to that in Ref.
\cite{Schmidt1} for $d\rightarrow \infty$
 since the boundary conditions used here are different.
\begin{figure}[th]
\begin{center}
\epsfig{file=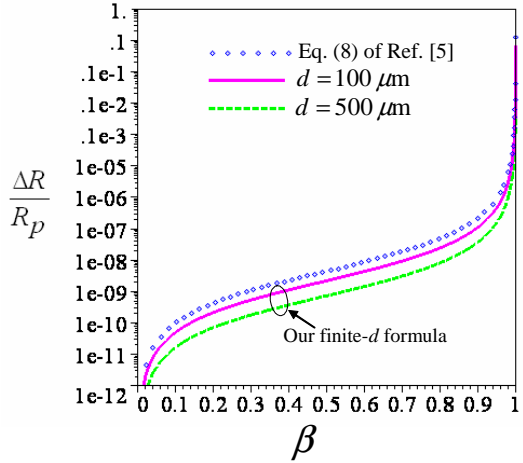,width=0.4\textwidth}
\end{center}
\caption{Color online. $\protect \Delta R/R_p$ as a function of $\protect \beta$
for FM/NMS/FM structures with finite FM layers $d$. The solid and
dashed curves correspond to the formula derived in the present
paper [Eqs. (\ref{resist}) and (\ref{anti})], while the diamond
curve is obtained from Eq. (8) of Ref. \onlinecite{Schmidt1}. Here
we use $\protect x_0=1 \mu$m, $\protect \lambda_D=10$ nm,
$\protect \lambda_N \rightarrow \infty$, $\protect d=100$ $\mu$m,
and $\protect \sigma^D=100\sigma^N$. The three curves do not
differ significantly for the parameters used.} \label{fig:fig2}
\end{figure}

\textit{Magnetoresistances.} For metallic FM/NMS/FM systems we
define the resistance change as usual, i.e., $\Delta
R=R_{ap}-R_{p}$. The full expression is too lengthy to be shown
here. However, in the physically relevant limit $\lambda _{D}\ll
d$ and $x_{0}\ll \lambda _{N}$ we find from Eqs. (\ref{resist})
and (\ref{anti})
\begin{equation}
\frac{\Delta R}{R_p}=\frac{\beta^2 (\sigma ^{N}\lambda
_{D}/x_{0}\sigma ^{D})^2 } {(1-\beta^2)[(1-\beta^2)+\frac{\sigma
^{N}\lambda _{D}}{x_{0}\sigma ^{D }} + (\frac{\sigma ^{N}\lambda
_{D}}{\lambda _{N}\sigma ^{D}})^2\frac{1}{(1-\beta^2)}]}.
\label{ourDelta}
\end{equation}
In deriving the above equation we have approximated $R_p$ in the
denominator by its dominant contribution $2x_0/\sigma^N$ from the
central NMS layer.  Again, our Eq. (\ref{ourDelta}) is different
from the corresponding one [Eq. (8)] in Ref. \cite{Schmidt1}.
However, as can be seen from comparison of Eq. (\ref{ourDelta})
and Eq. (8) of Ref. \cite{Schmidt1}, this deviation is not really
important for \textit{metallic} ferromagnets. Indeed, $\Delta
R/R_{p}$ which follows from our new equations practically
coincides with the quantity in Eq. (8) of Ref. \cite{Schmidt1} for
realistic values of the parameters when $\sigma ^{N}\lambda
_{D}/\sigma ^{D}\lambda _{N}\ll 1$, $\sigma ^{N}\lambda
_{D}/\sigma ^{D}x_0 \ll 1$, and for $\beta$'s not anomalously
close to unity.

In DMS/NMS/DMS systems, on the other hand, only the parallel
configuration is possible. In this case, we define the relevant
resistance change in the system with respect to the zero
magnetic-field resistance $R_0$, i.e., $\Delta R^{D}=R_p-R_0$. We
find
\begin{widetext}
\begin{equation}
\frac{\Delta R}{R_0}^D= \frac{\beta ^{2}}{(1-\beta
^{2})\frac{\sigma ^{D}}{\lambda _{D}}\coth (\frac{d}{\lambda
_{D}})+\frac{\sigma ^{N}}{\lambda _{N}}\coth (\frac{x_{0}}{\lambda
_{N}})} \frac{1}{\left( \frac{x_0}{\sigma^N} +
\frac{d}{\sigma^D}\right)}+ \frac{d (1/\sigma^D(H)-1/\sigma^D(0))}{\left( \frac{x_0}{\sigma^N} +
\frac{d}{\sigma^D}\right)}.
\label{deltar0}
\end{equation}
\end{widetext}
Note that $R_0$ is simply the combined NMS/DMS layer resistances
$2x_0/\sigma^N+2d/\sigma^D$, since no spin accumulation occurs for
zero magnetic field. In Eq. (\ref{deltar0}) we have indicated
explicitly the additional magnetoresistance arising from the DMS
(or FM) regions through $\sigma^D(H)$.

Note that $\Delta R^D/R_0$ goes to zero for either $d \rightarrow
\infty$ or $x_0 \rightarrow \infty$. The exact expression for
$\Delta R/R_p$ obtained from Eqs. (\ref{resist}) and (\ref{anti})
[not the approximate one in Eq. (\ref{ourDelta})] also vanishes in
these limits. Note that this vanishing of the resistance change is
symmetrical in $d$ and $x_0$ as expected. Earlier formulas in the
literature do not show this symmetry since they assume different
boundary conditions.

In contrast to metallic FM/NMS/FM systems, DMS/NMS/DMS structures
can have conductivities of the DMS and NMS regions which are
comparable. Besides, $\lambda_D$ can be of the same order as $d$
in some experiments. Experimental results in this case, e.g., the
\textit{d} dependence of the magnetoresistance effect, should be
interpreted with the use of Eq.(\ref{deltar0}). As we mentioned
earlier, the magnetoresistance effect observed in Ref. \cite{mol}
doubles with the DMS thickness. Let us find out whether our 1D
formula can explain this observed thickness dependence. Note that
the \textit{d}-dependence in Eq. (\ref{deltar0}) appears through
the $\coth(d/\lambda^D)$ term (spin accumulation effect). Since
$\sigma ^{D}=\sigma _{\uparrow}^{D}+\sigma_{\downarrow }^{D}$ can
in general be magnetic-field dependent, there is also a possible
\textit{d}-dependence arising from the second term which describes
the resistances of the DMS leads.
\par
In order to investigate in more detail the \textit{d}-dependence
in $\Delta R^D/R_0$, we need to know the magnetic-field
dependences of the quantities entering $\Delta R^D/R_0$ [Eq.
(\ref{deltar0})], particularly the magnetic field dependence of
the spin-dependent conductivities $\sigma^D_{\uparrow,
\downarrow}$. Unfortunately, these dependences are not known.
However, we can still gain some insight into the problem by
neglecting the magnetic field-dependence of the conductivities. In
this case, the second term in Eq.(\ref{deltar0}) is zero and we
can investigate the expected d-dependence arising from the
spin-accumulation effect alone. Note that the quantity $\beta$ is
null in the absence of a magnetic field and increases as a
function of it. Hence,  we plot $\Delta R^D/R_0$ as a function of
degree of spin-polarization $\beta$ for different values of $d$.
Figures 3(a), 3(b) show plots of $\Delta R^D/R_0$ vs $\beta$ for
different parameters. We have chosen the values of $d, x_0$ and
$\sigma^D/\sigma^N=1/3$ corresponding to the real experimental
situation of the Ref. \cite{mol}.  From these plots we can see
that even though the absolute value of the magnetoresistance is
comparable to the experimental value, the \textit{d}-dependence is
not as pronounced as in the experiment of Ref. \cite{mol}. It is
indeed true that the magnetoresistance effect is larger for
increasing $d$. However, this behavior only appears at values of
$\lambda_D$ comparable to the thickness $d$ and even for
$\lambda_D = 0.5$ $\mu$m $> d$, Fig. 3(b), the effect is way too
small to explain the experimental data. We conclude that the spin
accumulation effect by itself does not explain -- with realistic
parameters -- the observed \textit{d}-dependence of the data in
Ref. \cite{mol}.

\begin{figure}[th]
\begin{center}
\epsfig{file=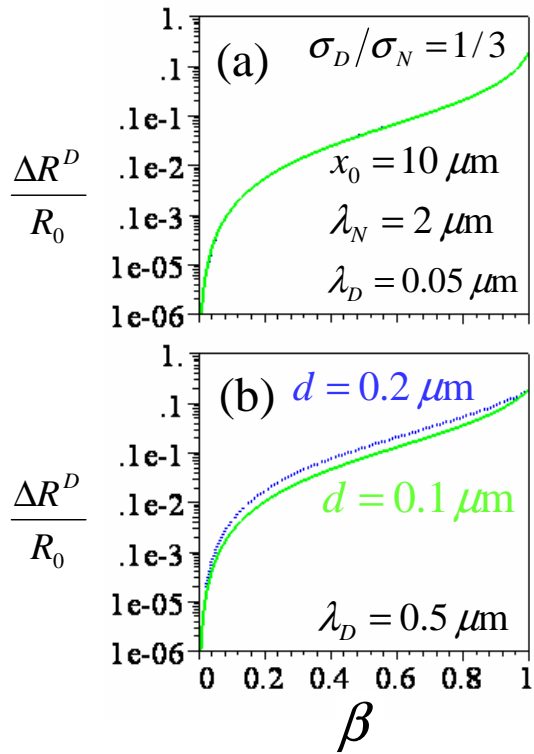,width=0.4\textwidth }
\end{center}
\caption{Color online. $\protect \Delta R^D/R_0$ vs $\protect \beta$ for two distinct DMS thicknesses and spin-flip lengths. Only the spin
accumulation part of the magnetoresistance, i.e., the first term
in Eq. (\ref{deltar0}), is considered here. For $d\gg\lambda_D$
the magnetoresistance does not show any dependence on $d$ (a). On
the other hand, for $\lambda_D$ comparable to $d$ the
magnetoresistance (b) increases slightly with $d$; this effect is
too small to explain the $d$-dependence of the magnetoresistance
in Ref. \cite{mol}. } \label{fig:fig3}
\end{figure}

Let us mention here the possibility to explain the experiment by
the magnetoresistance of the leads, see the second term in Eq.
(\ref{deltar0}). Control experiments on a layer of the DMS
injector material patterned into a Hall geometry showed $\sim 1
\%$ negative MR \cite{mol}. However, the transport problem we deal with
here can be shown to reduce to that of a 1D system
\cite{Khaetskii}. Note that in the case when the magnetic field
dependence of $\sigma^D$ is well pronounced and for realistic
values $x_0 \gg d \gg \lambda_D$ and for $\beta \simeq 1$ the
first term in  Eq. (\ref{deltar0}) is estimated as $\lambda_D
/\sigma^D$ while the second term is of the order of $d/\sigma^D$.
Thus in the case when the magnetoresistance of the leads is not
zero the $d$-dependence associated with it can be more pronounced
than the one associated with the spin accumulation effect.

\textit{Some limiting cases.} From the general equations we
derived above, we can obtain expressions for some interesting
limiting cases. If $\sigma _{\downarrow }^{D}\rightarrow 0$, i.e.
$\beta \rightarrow 1$, we find from Eq. (\ref{resist})
\begin{equation}
R_{p}=\frac{2x_{0}}{\sigma ^{N}}+\frac{2d}{\sigma _{\uparrow }^{D}}+2\frac{%
\lambda _{N}}{\sigma ^{N}}\tanh (x_{0}/\lambda _{N}).
\end{equation}
If, in addition, $\lambda _{N}\rightarrow \infty $, then
\begin{equation*}
R_{p}=\frac{2x_{0}}{\sigma ^{N}}+\frac{2d}{\sigma _{\uparrow }^{D}}+\frac{%
2x_{0}}{\sigma ^{N}}.
\end{equation*}%
Hence the resistance of the central part (NMS) doubles as expected. Note
that the resistance of the DMS part does not necessarily double.

\textit{Current spin polarization.} We can also easily determine
some other relevant quantities in our system. For example, the
degree of spin polarization $\alpha(x)$ of the current at the
DMS/NMS (or FM/NMS) interface ($x=x_0$) is equal to
\begin{equation}
\alpha (x_0) =\frac{j_{\uparrow }-j_{\downarrow }}{j_{\uparrow }+j_{\downarrow }}=%
\frac{\beta }{1+\frac{\sigma ^{D}}{\sigma ^{N}}(1-\beta
^{2})\frac{\lambda _{N}\tanh (x_{0}/\lambda
_{N})}{\lambda_{D}\tanh (d/\lambda_{D})}}, \label{alpha}
\end{equation}
for parallel alignment. This result coincides exactly with Eq. (7)
of Ref. \cite{Schmidt1} in the limit $x_{0}\ll \lambda
_{N},\/\lambda _{D}\ll d$. For metallic ferromagnetic contacts we
can have antiparallel alignment as well. Note that $\alpha(x)$ is
an even (odd) function of $x$ for parallel (antiparallel)
alignment.

For completeness, we give below some other expressions for
$\alpha(x)$. At $x=\pm(x_0+d)$ (even function) we have
\begin{equation}
\alpha(x=-x_0-d) = \beta -\frac{\beta}{\cosh(d/\lambda_{D})[1+
\frac{\sigma ^{N}}{\sigma ^{D}}\frac{1}{(1-\beta
^{2})}\frac{\lambda _{D}\tanh
(d/\lambda_{D})}{\lambda_{N}\tanh(x_{0}/\lambda _{N})}]}.
\label{alpha1}
\end{equation}
Note that for $d\gg \lambda_D$ $\alpha(x=-x_0-d)$ approaches the
maximal possible value $\beta$.  In addition
\begin{equation}
\alpha(x=0)=\alpha(x=-x_0)\frac{1}{\cosh(x_0/\lambda_{N})}.
\label{alpha2}
\end{equation}
Thus in the middle of the sample the spin current polarization is
exponentially suppressed. In the case of antiparallel
configuration the corresponding values of $\alpha$ at $x=-x_0$ and
$x=-x_0-d$ can be obtained from formulas (\ref{alpha}) and
(\ref{alpha1}), respectively, by replacing $\tanh (x_{0}/\lambda
_{N})$ for $\coth(x_{0}/\lambda _{N})$.

In conclusion, we have derived the expressions for the resistances
of FM/NMS/FM (or DMS/NMS/DMS) structures using physical boundary
conditions at the corresponding interfaces. We have considered
magnetic layers (DMS or FM) of finite length $d$. When the
magnetoresistance of the contacts is negligible (i.e., when
$\sigma_D$ is not magnetic field dependent), we find that the $d$
dependence of the spin-accumulation contribution to the
magnetoresistance is not enough to explain the experimental
results of Ref. \cite{mol}. On the other hand, in systems where
$\sigma_D$ is magnetic field dependent our formulas allow for a
sizable $d$ dependence arising from the magnetoresistance of
contacts. Interestingly, for realistic parameters $x_0\gg d \gg
\lambda_D$ and with $\beta \simeq 1$ the magnetoresistance of the
contacts is larger than that from the spin accumulation effect.

This work was supported by the Swiss NSF, NCCR Nanoscience Basel, DARPA,
ARO, ONR and the DARPA Spins program.


\begin{references}

\bibitem{wolf}
S.~A. Wolf, D.~D. Awschalom, R.~A. Buhrman, J.~M. Daughton, S. von
Molnar, M.~L. Roukes, A.~Y. Chtchelkanova, and D.~M. Treger,
Science {\bf 294}, 1488 (2001).

\bibitem{als}
{\em Semiconductor Spintronics and Quantum Computation}, eds.
D.~D. Awschalom, D. Loss, and N. Samarth, Springer, Berlin, 2002.


\bibitem{Schmidt}
G. Schmidt and L.W. Molenkamp, Semicond. Sci. Technol. {\bf 17}, 310 (2002).


\bibitem{datta}
 S. Datta and B. Das, Appl.\ Phys.\ Lett. \textbf{56}, 665
(1990). J. C. Egues, G. Burkard, and D. Loss, Appl. Phys. Lett.
\textbf{82}, 2658 (2003); J. Schliemann, J. C. Egues, and D. Loss,
Phys. Rev. Lett \textbf{90}, 146801 (2003).

\bibitem{Schmidt1}
G. Schmidt, D. Ferrand, L.W. Molenkamp, A.T. Filip, and B.J. van Wees, Phys. Rev.B {\bf 62}, R4790 (2000).

\bibitem{rashba}
 E. I. Rashba, Phys. Rev. B {\bf 62}, R16267
(2000); A. Fert and H. Jaffr$\grave{\rm e}$s, Phys. Rev. B {\bf
64} 184420 (2001).

\bibitem{motsnyi}
V. F. Motsnyi, J. De Boeck,
J. Das, W. Van Roy, G. Borghs, E. Goovaerts, and V. I. Safarov,
Appl. Phys. Lett \textbf{81}, 265 (2002).

\bibitem{egues}
J. C. Egues, Phys.\ Rev.\ Lett.\ \textbf{80}, 4578
(1998).

\bibitem{spin-pol}
R.\ Fiederling, M. Keim, G. Reuscher, W. Ossau, G. Schmidt, A. Waag, and 
L.W. Molenkamp, Nature \textbf{402}, 787
(1999); Y. Ohno, D.K. Young, B. Beschoten, F. Matsukura, H. Ohno, and D.D. Awschalom, Nature \textbf{402}, 790 (1999).

\bibitem{mol}
G. Schmidt, G. Richter, P. Grabs, C. Gould, D. Ferrand, and L.W. Molenkamp, Phys. Rev. Lett. {\bf 87}, 227203 (2001).

\bibitem{Khaetskii}
It should be mentioned that despite the 2D geometry of the real
experiment \cite{mol}, the transport problem reduces to that of a
1D system. This result follows from the specific current density
distribution established by minimizing the Joule heat (A.
Khaetskii, unpublished). In a real experimental situation the DMS
leads lie on the top of the NMS layer (at $|x| > x_0$), thus the
current lines are bent and the current enters the metallic
contacts in the direction perpendicular to the original
$x$-direction. An \textit{exact} analytical treatment of this
geometry, shows that the current which enters or leaves the NMS
layer flows only near the corners of the NMS/DMS interfaces. It
turns out that the relevant width $\Delta x$ of the DMS layer
where transport occurs is of the order of the NMS layer height.
The current density exponentially decays for $|x|\geq (x_0 +
\Delta x)$. Therefore we have essentially a ``bent'' 1D geometry.

\bibitem{formulas}
See for instance, S. Hershfield and H. L. Zhao, Phys. Rev. B
\textbf{56}, 3296 (1997), E. I. Rashba, Eur. Phys. J. B
\textbf{29}, 513 (2002) and Z. G. Yu and M. E. Flatt\'{e}, Phys.
Rev. B \textbf{66}, 235302 (2002).

\bibitem{vanson}
P. C. van Son, H. van Kempen, and P. Wyder, Phys. Rev. Lett.
\textbf{58}, 2271 (1997).

\bibitem{correction}
Note that Eq.(1) of Ref. \cite{mol} contains a misprint. In the
numerator of this equation, instead of the first power of $\beta$
it should be $\beta^2$. The correct version of this formula can be
found in: G. Schmidt and L.W. Molenkamp, in {\it Semiconductor
Spintronics and Quantum Computing}, (Springer-Verlag, Berlin
Heidelberg 2002, D.D. Awschalom, D. Loss, N. Samarth, Eds.). Note,
however, that our Eq. (\ref{resist}) contains totally different
$d$ and $x_0$ dependences as compared to Eq. (1) of Ref.
\cite{mol}.




\end{references}
\end{document}